\documentclass[aps,prb,twocolumn]{revtex4} 

\usepackage{graphicx}
\usepackage{dcolumn}
\usepackage{bm}
\usepackage{amsmath}  

\newcommand{\comment}[1]{}

\begin{document}

\title{
The molecular nature of superfluidity:
Viscosity of helium from quantum stochastic molecular dynamics simulations
over real trajectories}


\author{Phil Attard}
\affiliation{ {\tt phil.attard1@gmail.com}  July 2024--Feb.\ 2025}
\noindent {\tt  Projects/QSM24/SFVisco/ViscoSF.tex}


\begin{abstract}
Using quantum equations of motion for interacting bosons,
stochastic molecular dynamics simulations with quantized momenta
are performed for Lennard-Jones helium-4.
The viscosity of the quantum liquid
is significantly less than that of the classical liquid,
being almost 5 times smaller at the lowest temperature studied.
The classical and quantum liquids are identical
except for Bose-Einstein condensation,
which pinpoints the molecular mechanism for superfluidity.
The results rely on the existence of stochastic
but real particle trajectories,
which has implications for the interpretation of quantum mechanics.
\end{abstract}

\pacs{}

\maketitle

%
\section{Introduction}
\setcounter{equation}{0} \setcounter{subsubsection}{0}
\renewcommand{\theequation}{\arabic{section}.\arabic{equation}}
%

\begin{equation}
\mbox{Shut up and calculate}
\end{equation}
is an aphorism  apparently due to Mermin,
although some say Feynman (Mermin 1989, 2004).
The sentiment,
which is widespread and likely predates the specific phrase,
suggests that it is a waste of time to speculate
about the interpretation of quantum mechanics
since everyone agrees upon the fundamental equations.
It acknowledges
the spookiness of quantum non-locality
and the jittery state of Schr\"odinger's cat,
but it insists that the physical meaning of these has no bearing
on the application of the quantum laws and equations,
which themselves are unambiguous.

It is certainly true
that any theory or calculation is bound by the mathematical rules.
Nevertheless I think that it goes too far to say that
the physical interpretation of those rules is irrelevant
or that the discussion of fundamental quantum concepts is mere sophistry.
Once one goes beyond the highly idealized undergraduate textbook examples
to work on real world problems,
it is necessary to introduce approximations
into the fundamental quantum equations.
Such approximations may involve neglecting particular classes of terms
while resumming others,
defining parameters and taking small or large asymptotic limits,
choosing specific functions for expansion series,
imposing particular boundary conditions, etc.
The approach chosen
depends not just upon  the physical characteristics
of the problem at hand
but also upon the interpretation of quantum mechanics
that leads to an understanding of what is important
and what is negligible,
what is doable and what is forbidden.
It is often the case that the real world application
is so far removed from the fundamental quantum equations
that the results of an early decision for the research direction
cannot be fully anticipated or easily undone.
Different interpretations can lead to widely divergent theories
due to the sensitivity to the initial beliefs, if you will.

Let me give a concrete example
that is directly relevant to the present paper
on computing the viscosity of helium
in the superfluid regime.
The Copenhagen interpretation of quantum mechanics
holds that the world is not objectively real
and that it only comes into existence when it is measured or observed.
More specifically,
particles only possess position or momentum at the time of measurement,
and that only one of these can be measured at a time.
Therefore, it is said,
a particle does not  possess simultaneously position and momentum.
The corollary of this is that particles
cannot follow a path from one position-momentum point to another,
which is to say that particle trajectories do not exist.

Obviously any scientist who wishes to understand superfluidity
at the molecular level
and who believes in the Copenhagen interpretation of quantum mechanics
would never consider developing a theory or approximation
that is based upon real particles with actual positions and momenta
following actual trajectories in time.
This example illustrates how a particular interpretation
of quantum mechanics can proscribe from the start
the theories or approximations that are even considered,
let alone explored.

The results in this paper are based on real particles
with simultaneously specified positions and momenta,
and on real molecular trajectories in time.
These obviously contradict the Copenhagen interpretation
of quantum mechanics.
But do they contradict the equations of quantum mechanics?
Obviously I argue not, as I now briefly explain.

It is certainly true that the position and momentum operators
do not commute and that Heisenberg's uncertainty principle
bounds the product of the variance of the expectation values
of the  position and momentum operators.
These are indisputable mathematical facts.
Anything beyond these is a matter of interpretation,
and highly questionable interpretation at that.
For example,
the assertion that an expectation value is a measurement is dubious;
there are entire journals devoted to the quantum theory of measurement
and the only thing that the various authors agree upon
is that a measurement is not simply an expectation value.
Further,
it is not at all clear that the lack of commutativity of the
position and momentum operators
implies the Copenhagen interpretation
that a particle cannot possess simultaneously
a position and a momentum.
For a counter-example, see the de Boglie-Bohm pilot wave theory,
which reproduces all of the known results of quantum mechanics
(Bohm 1952, de Broglie 1928, Goldstein 2024).

The approach used here is predicated on the interpretation
that it is the momentum eigenvalue that gives the momentum of a particle,
and that a momentum eigenfunction at a particular position
should be interpreted
as a complex number that is associated with  the simultaneous specification
of the position and momentum of the particle.
In a way that will be made clear,
the state of the system
is the product of single-particle momentum eigenfunctions of the subsystem,
and the subsystem evolves in time by following
a trajectory through classical phase space
as given by the Schr\"odinger equation
applied to the momentum eigenfunctions
and taking into account the interactions with the environment.
It is essential to this approach that the subsystem of interest be open
and that it can exchange energy and momentum with its environment.
(The total system consists of the subsystem
and the reservoir or environment.)
It is also essential that the symmetrization
of the wave function be explicitly accounted for.

At the end of the day,
the present interpretation and the consequent approximations
that are made should be judged by their physical plausibility
and by the results that they produce.
One cannot really maintain that the interpretation of quantum mechanics
is irrelevant to the real world
if the same starting equations combined with different interpretations
lead to different quantitative descriptions of that world.
The present approach gives a quantitative estimate
of the shear viscosity of superfluid helium
that includes molecular interactions.
Apart from related work by the present author (Attard 2023b, 2025),
these are the first such molecular-level results
for the superfluid viscosity.
It is therefore reasonable to conclude that the proximal impediment
to the molecular understanding
and quantitative description of superfluidity
has been the Copenhagen interpretation of quantum mechanics.

%
\section{Analysis}
\setcounter{equation}{0} \setcounter{subsubsection}{0}
\renewcommand{\theequation}{\arabic{section}.\arabic{equation}}
%

\subsection{Hamiltonian Dynamics}

Quantum statistical mechanics
may be formulated in classical phase space (Attard 2018, 2021, 2023a).
This exact transformation relies upon decoherence
due to entanglement with the reservoir or environment.
This configuration picture
invokes the  momentum eigenfunction
$\phi_{\bf p}({\bf q})
= V^{-N/2} e^{-{\bf p} \cdot  {\bf q}/\mathrm{i}\hbar}$
(Merzbacher  1970, Messiah 1961).
Here for $N$ particles the momentum configuration is
${\bf p} = \{{\bf p}_1,{\bf p}_2,\ldots,{\bf p}_N\}$,
and the position configuration is
${\bf q} = \{{\bf q}_1,{\bf q}_2,\ldots,{\bf q}_N\}$.
The momenta are quantized,
with momentum state spacing $\Delta_p=2\pi\hbar/L$,
with $V=L^3$ being the volume of the cubic subsystem,
and $\hbar=1.05\times10^{-34}$J\,s being Planck's constant
divided by $2\pi$.
The spacing between momentum states goes to zero
in the thermodynamic limit.
Note that a point in classical phase space,
${\bm \Gamma} \equiv \{{\bf q},{\bf p}\}$,
has the interpretation of a specific configuration of bosons
at these positions with these (quantized) momenta,
and it has associated with it the complex number $\phi_{\bf p}({\bf q})$.

For bosons the normalized symmetrized momentum eigenfunction is
\begin{equation}
\phi^+_{\bf p}({\bf q}) =
\frac{1}{\sqrt{N!\chi_{\bf p}^+}} \sum_{\hat{\mathrm P}}
\phi_{\hat{\mathrm P}{\bf p}}({\bf q}) ,
\end{equation}
$\hat{\mathrm P}$ being the permutation operator.
The symmetrization factor  is
\begin{eqnarray}
\chi_{\bf p}^+
& = &
\sum_{\hat{\mathrm P}}
\langle \phi_{\bf p}  |
\phi_{\hat{\mathrm P}{\bf p}}  \rangle
\nonumber \\ & = &
\sum_{\hat{\mathrm P}}
\delta_{{\bf p},\hat{\mathrm P}{\bf p}}
\nonumber \\ & = &
\prod_{\bf a} N_{\bf a}({\bf p})! .
\end{eqnarray}
Here and throughout the occupancy
of the single particle momentum state ${\bf a}$
is $N_{\bf a} = \sum_{j=1}^N \delta_{{\bf p}_j,{\bf a}}$.
This is what ultimately drives Bose-Einstein condensation (Attard 2025).

The Born probability associated with a point in classical phase space
for the subsystem in a symmetrized decoherent momentum state is
(Attard 2025 Eq.~(5.67))
\begin{eqnarray}
\lefteqn{
\phi^+_{\bf p}({\bf q})^*\,\phi^+_{\bf p}({\bf q})
} \nonumber \\
& = &
\frac{1}{V^{N}N!\chi^+_{\bf p}}
\sum_{\hat{\mathrm P}',\hat{\mathrm P}''}
e^{-(\hat{\mathrm P}'{\bf p}-\hat{\mathrm P}''{\bf p})\cdot{\bf q}
/\mathrm{i}\hbar}
\nonumber \\ & \approx &
\frac{1}{V^{N}N!\chi^+_{\bf p}}
\sum_{\hat{\mathrm P}',\hat{\mathrm P}''}
\!\!\!^{(\hat{\mathrm P}'{\bf p} \approx \hat{\mathrm P}''{\bf p})}
e^{-(\hat{\mathrm P}'{\bf p}-\hat{\mathrm P}''{\bf p})\cdot{\bf q}
/\mathrm{i}\hbar} .
\end{eqnarray}
The reason for neglecting the terms
involving permutations of bosons with dissimilar momenta
is that these are more or less randomly and uniformly distributed
on the unit circle in the complex plane,
and so they add up to zero.
This is particularly the case when one considers
that small changes in the positions
may lead to wildly different exponents for any such permutations.
The sum that is retained involves only permutations between bosons
in the same, or nearly the same, momentum state,
in which case the exponent is zero, or close to zero,
even for small changes in positions.
The number of permutations between bosons
in exactly the same momentum states in the double sum is
$\sum_{\hat{\mathrm P}',\hat{\mathrm P}''}
\!^{(\hat{\mathrm P}'{\bf p} = \hat{\mathrm P}''{\bf p})}
= N! \prod_{\bf a} N_{\bf a}({\bf p})!
= N!\chi^+_{\bf p} $.
Arguably these are the ones that dominate,
particularly on the low side of the $\lambda$-transition,
which means that we may take
$\phi^+_{\bf p}({\bf q})=\phi_{\bf p}({\bf q})$.

An open system is decoherent
(Attard 2018, 2021, Joos and Zeh 1985, Schlosshauer 2005, Zurek 1991).
Decoherence means that the only allowed permutations
must satisfy $\hat{\mathrm P}{\bf p} ={\bf p}$.
Otherwise the symmetrized momentum eigenfunction,
$\phi_{\bf p}^+({\bf q})$,
would be a superposition of states.

There is a decoherence time $\tau_\mathrm{mix}$
(Caldeira  and Leggett  1983,
Schlosshauer 2005, Zurek \emph{et al.}\ 2003).
This likely decreases with increasing distance
between permuted momentum states.

Schr\"odinger's equation
for the time evolution of the momentum eigenfunction
in a decoherent system for a small time step gives
(Attard 2023d, 2025),
\begin{equation}
[\hat{\mathrm I}
 + (\tau/\mathrm{i}\hbar)\hat{\cal H}({\bf q})] \phi_{\bf p}({\bf q})
= \phi_{{\bf p}'}({\bf q}').
\end{equation}
Time reversibility and continuity imply that
\begin{eqnarray}
{\bf q}'
& = &
{\bf q} + \tau \nabla_p {\cal H}({\bf q},{\bf p}),
\nonumber \\
\mbox{ and }
{\bf p}'
& = &  {\bf p} - \tau \nabla_q {\cal H}({\bf q},{\bf p}).
\end{eqnarray}
These are Hamilton's classical equations of motion.
The second says that
a boson's momentum evolves according to the classical force
acting on it,
$\dot {\bf p}_j
= -  \nabla_{q,j} {\cal H}({\bf q},{\bf p})
= {\bf f}_j$.
Here it is implicitly assumed that the spacing between momentum states
is small enough to allow the continuum approximation.

It should be emphasized
that there are two sources of decoherence:
there is the internal decoherence that was discussed above
on the basis of the Born probability,
and there is the external decoherence due to the entanglement
of the open subsystem with the environment or reservoir.
Both mechanisms dictate that the subsystem must be in a pure state
where each boson has a position and a momentum.
Allowed permutations are between bosons in the same momentum state.
Those permutations which would correspond to the superposition
of different momentum configurations are suppressed.


\subsection{Newton's Second Law: The Condensed Version}
\label{Sec:EoM-Cond}

For a momentum configuration ${\bf p}$ in the condensed regime,
the single-particle quantized momentum state ${\bf a}$
is occupied by $N_{\bf a} = \sum_{j=1}^N \delta_{{\bf p}_j,{\bf a}}$
bosons.
Let the $n_A$ bosons in a specific subset $A \in {\bf a}$ be
$A = \{ j_1,j_2,\ldots,j_{n_A} \}$,
with ${\bf p}_{j_k} = {\bf a}$.
The momentum eigenfunction for this subset
is $\phi_{{\bf p}^{n_A}}({\bf q}^{n_A})$.
Let $\hat{\mathrm P}_{\!A}$ be one of the $n_A!$ permutations
amongst the bosons in the specific subset $A$.
Obviously the momentum configuration is unchanged
by such a permutation,
$\hat{\mathrm P}_{\!A} {\bf p}^{n_A} = {\bf p}^{n_A}
={\bf a}^{n_A}$.
The Schr\"odinger time propagator for this subset yields
\begin{eqnarray}
\left[\hat{\mathrm I}
+ \frac{\tau}{\mathrm{i}\hbar}
\hat{\cal H}({\bf q}^{n_A}) \right]
\phi_{{\bf p}^{n_A}}({\bf q}^{n_A})
& = &
\phi_{ {\bf p}'^{n_A}}( {\bf q}'^{n_A})
 \\ & = & \nonumber
\frac{1}{n_A!} \sum_{\hat{\mathrm P}_{\!A}}
\phi_{\hat{\mathrm P}_{\!A}{\bf p}'^{n_A}}( {\bf q}'^{n_A}) ,
\end{eqnarray}
since the left hand side is unchanged by the sum over permutations.
This suggests that the subset moves rigidly,
which is to say $ {\bf p}_{j_k}' = {\bf a}'_A$
for ${j_k} \in A$,
since this ensures that
$\hat{\mathrm P}_{\!A}{\bf p}'^{n_A} = {\bf p}'^{n_A}$.

The \emph{classical} equations of motion become
\begin{eqnarray}
{\bf p}'^{n_A}
& = &
\frac{1}{n_A!} \sum_{\hat{\mathrm P}_{\!A}}
\hat{\mathrm P}_{\!A} {\bf p}'^{n_A}
\nonumber \\ & = &
{\bf p}^{n_A}
+
\frac{\tau}{n_A!} \sum_{\hat{\mathrm P}_{\!A}}
\hat{\mathrm P}_{\!A} {\bf f}^{n_A}
\nonumber \\ & = &
{\bf p}^{n_A} + \tau {\bf F}_A^{n_A} .
\end{eqnarray}
It is obvious that symmetrizing the force by permutation over the subset
is equivalent to the average non-local force per boson for the subset,
$
{\bf F}_A \equiv
(n_A!)^{-1} \sum_{\hat{\mathrm P}_{\!A}}
\big\{ \hat{\mathrm P}_{\!A} {\bf f}^{n_A} \big\}_{j_k}
=  n_A^{-1} \sum_{j\in A} {\bf f}_j$.

There are 
$^{N_{\bf a}}C_{n_A} \equiv
N_{\bf a}!/(N_{\bf a}-n_A)!n_A!$
ways of allocating labeled bosons
from the momentum state ${\bf a}$ with $N_{\bf a}$ bosons
to a subset $A$ with $n_A$ bosons.
Label these $n=1,2,\ldots,^{N_{\bf a}}\!C_{n_A}$.
Each one of these has an indistinguishable initial configuration,
${\bf a}^{n_A}$,
but a different average non-local force
${\bf F}_{A_n}$ and hence a different evolved configuration
$ {\bf p}'^{n_A}_n$.
The evolved configuration is a superposition of these.
But as an open quantum system,
it must collapse into a single configuration.
The specific configuration occurs with probability
$1/\,^{N_{\bf a}}C_{n_A} $, 
which means that
the quantum equation of motion
for the evolution of the specific subset of $n_A$ bosons
in the momentum state ${\bf a}$ is
\begin{equation} \label{Eq:pnap}
{\bf p}'^{n_A}
=
{\bf p}^{n_A}
+
\tau \frac{(N_{\bf a}-n_A)!n_A!}{N_{\bf a}!}  {\bf F}_{A}^{n_A} ,
\end{equation}
where the $A$ refers to a specific labeled subset
of the $^{N_{\bf a}}C_{n_A}$ possible subsets.

We can confirm that this is correct and gain some insight
into the physical process by looking at energy conservation.
If the momentum state is divided into the subset $A$ with $n_A$ bosons,
and $A'$ with $n_{A'}=N_{\bf a}-n_A$ bosons,
then the respective forces per boson are ${\bf F}_A$ and ${\bf F}_{A'}$.
The total force,
$n_{A} {\bf F}_{A_n} + n_{A'}{\bf F}_{A'_n} = \sum_{j\in a} {\bf f}_j$,
is the same for each of the $^{N_{\bf a}}C_{n_A}$ possible $A_n$.
The change in total potential energy over the time step
for the bosons in ${\bf a}$ is
$\Delta U = -\tau \sum_{j\in a} {\bf f}_j \cdot {\bf a}  /m$
because there is just one position configuration.
With the above condensed equation of motion
the change in total kinetic energy for the superposed momentum states is
\begin{eqnarray}
\Delta{\cal K}
&=&
\frac{1}{m} \sum_{n=1}^{^{N_{\bf a}}\!C_{n_A}}
\Big\{ {\bf a}^{n_A} \cdot [{\bf p}'^{n_A}_n - {\bf a}^{n_A}  ]
\nonumber \\ && \mbox{ }
+ {\bf a}^{n_{A'}} \cdot
[{\bf p}'^{n_{A'}}_n - {\bf a}^{n_{A'}} ] \Big\}
\nonumber \\ & = &
\frac{\tau}{m}  \frac{n_{A'}!n_A!}{N_{\bf a}!}
\sum_{n=1}^{^{N_{\bf a}}\!C_{n_A}}
\Big\{ n_A {\bf a} \cdot  {\bf F}_{A_n}
+ n_{A'} {\bf a} \cdot {\bf F}_{A_n'} \Big\}
\nonumber \\ & = &
\frac{\tau}{m} \sum_{j\in a} {\bf a} \cdot {\bf f}_j.
\end{eqnarray}
This cancels with the change in potential energy.
Hence the superposed evolved momentum states
maintain constant energy for the bosons originally
in the momentum state ${\bf a}$.
It is essential for this result
that the force on the subsets $A$ and $A'$
be reduced by the binomial coefficient.

The result for the evolution of the momentum of a subset of bosons,
Eq.~(\ref{Eq:pnap}),
says that Newton's classical rate of change of momentum
is shared amongst the $^{N_{\bf a}}C_{n_A}$ superposed states,
which in total conserves energy.
But since only one momentum configuration survives,
we conclude that kinetic energy is lost to the environment via entanglement.
This result explains physically the conditional transition probability
derived next.

\subsection{Adiabatic Stochastic Transition} \label{Sec:AdTrans}

The change in position of the bosons over a time step $\tau$ is
deterministic,
\begin{equation}
{\bf q}(t+\tau) = {\bf q}(t) + \frac{\tau}{m} {\bf p}(t).
\end{equation}
We now explicitly take the momenta to be quantized,
with ${\bf p}$ being a $3N$-dimensional vector
integer multiple of $ \Delta_p$.

As in the preceding \S\ref{Sec:EoM-Cond},
we consider the transition of a
subset $A$ of bosons in the momentum state ${\bf a}$
that moves rigidly
according to the shared non-local force
acting on the subset.
The size distribution for the subsets will be derived
in the following \S\ref{Sec:gna}.

The present rigid subset model
allows the momentum state to break up
and the bosons ties to rearrange according to the state that they
occupy after each time step or even after each successful transition.
This shifts from the occupancy picture
to the configuration picture,
and therefore changes in occupation entropy must be taken into account.
The configuration probability density is (Attard 2025)
\begin{equation}
\wp({\bf \Gamma}) =
\frac{1}{Z} e^{-\beta {\cal K}({\bf p})}  e^{-\beta U({\bf q})}
\prod_{\bf a} N_{\bf a}! ,
\end{equation}
where $\beta=1/k_\mathrm{B}T$ is the inverse temperature,
${\cal K}({\bf p})$ is the kinetic energy,
$U({\bf q})$ is the potential energy,
the commutation function has been neglected,
and the occupancy of the momentum state ${\bf a}$ is
$N_{\bf a} = \sum_{j=1}^N \delta_{{\bf p}_j,{\bf a}}$.
Also, a point in quantized phase space
is ${\bf \Gamma} = \{{\bf q},{\bf p}\}$,
and the conjugate point with momenta reversed is
${\bf \Gamma}^\dag = \{{\bf q},-{\bf p}\}$.

Consider a specific subset $A \in {\bf a}$ of $n_A$ bosons.
The shared non-local force per boson in the subset is
${\bf F}_A = n_A^{-1} \sum_{j\in A} {\bf f}_j$.
We seek the conditional transition probability
of this subset to the neighboring momentum state
in the direction of the $\alpha$ component
of the force on the subset,
namely from ${\bf a}$ to
${\bf a}'_\alpha = {\bf a}
+ \mbox{sign}(\tau F_{A,\alpha})\Delta_p \widehat{\bf x}_\alpha$.
Microscopic reversibility gives
the ratio of conditional transition probabilities,
\begin{eqnarray}
\frac{ \wp({\bf \Gamma}'|{\bf \Gamma};\tau)
}{ \wp({\bf \Gamma}^\dag|{\bf \Gamma}'^\dag;\tau) }
& = &
\frac{\wp({\bf \Gamma}')}{\wp({\bf \Gamma})}
\nonumber \\ & = &
e^{(-n_A\beta /2m) [ a'^2-a^2] }
e^{({ n_A\beta \tau }/{m}) {\bf F}_{A} \cdot {\bf a} }
\nonumber \\ && \mbox{ } \times
\frac{(N_{{\bf a}'}+n_A)! (N_{{\bf a}}-n_A)!}{N_{{\bf a}'}! N_{{\bf a}}!}.
\end{eqnarray}
(This does not include the size probability distribution of the subsets;
see \S\ref{Sec:gna}.)
By inspection,
this is satisfied by a conditional transition probability
with the general form
\begin{eqnarray} \label{Eq:wploop}
\lefteqn{
\wp({\bf a}'_\alpha|{\bf a},n_A,N_{\bf a}, N_{{\bf a}'_\alpha})
}  \\
& = &
\lambda_{A,\alpha}
g(n_A)
\left(\frac{(N_{{\bf a}'_\alpha}+n_A)!}{N_{{\bf a}'_\alpha}!}\right)^{1-r}
\left(\frac{ (N_{{\bf a}}-n_A)! }{ N_{{\bf a}}! }\right)^r
\nonumber \\ && \mbox{ } \times \nonumber
e^{(-n_A\beta /2m) [ a'^2_\alpha-a_\alpha^2]/2 }
e^{({ n_A\beta \tau }/{m}) {F}_{A,\alpha} {a}_\alpha /2}
\nonumber \\ & \Rightarrow &
\lambda_{A,\alpha}
\frac{ n_A! (N_{{\bf a}}-n_A)! }{ N_{{\bf a}}! }
\nonumber \\ && \mbox{ } \times \nonumber
\left\{ 1
- \frac{n_A\beta \Delta_p}{2m} \mbox{sign}(\tau {F}_{A,\alpha}) a_\alpha
+ \frac{ \beta \tau a_\alpha }{2m} \sum_{j\in A} {f}_{j,\alpha}  \right\} .
\end{eqnarray}
The exponential of the change in energy has been linearized here,
although this is unnecessary
and might in fact be ill-advised for larger subsets.
In the particular solution that is the final equality,
$r=1$ has been chosen. 
Likewise,
$g(n_A)= n_A! $ has been chosen
(see \S\ref{Sec:EoM-Cond};
but similar values for the viscosity
result from $g(n_A)=\delta_{n_A,1}$ (see \S\ref{Sec:gna}),
$g(n_A)= 1$,
and the rigid state model $g(n_A)=\delta_{n_A,N_{\bf a}}$).
This gives
the known conditional transition probabilities
for a rigid state transition, $n_A=N_{\bf a}$.

\comment{ 
The algorithm is implemented
with an adiabatic transition attempted sequentially
for each component of momentum of each subset in the system
at each time step using the shared non-local force for subset,
${F}_{A,\alpha}$
(similar values for the viscosity result from using instead
${F}_{{\bf a},\alpha}$).
The occupancies were updated at the end of the time step,
although similar values for the viscosity result
when they are updated after each  successful transition.
The probability of a successful transition is low
in the limit $\tau \to 0$.
} 

With this conditional transition probability,
the average rate of change of momentum
in the direction $\alpha$
for the subset $ A \in {\bf a}$ to leading order is
\begin{equation}
\left\langle \dot p_{A}^0 \right\rangle
=
\frac{n_A \Delta_p}{\tau }
\mbox{sign}(\tau F_{A,\alpha})
\lambda_{A,\alpha}
\frac{n_A! (N_{{\bf a}}-n_A)! }{ N_{{\bf a}}! }.
\end{equation}
The classical regime is defined as $N_{{\bf a}} = n_A = 1$,
and in order to satisfy  Newton's second law of motion in this case
we must have
\begin{equation}  \label{Eq:lambda}
\lambda_{A,\alpha}
\equiv
\frac{|\tau F_{A,\alpha}|}{\Delta_p} .
\end{equation}
In the classical regime ${\bf F}_{\bf a} = {\bf F}_{A} = {\bf f}_j$
for $j \in A \in {\bf a}$.

With this result for $\lambda_{A,\alpha}$
and the conditional transition probability,
Newton's second law is \emph{not} satisfied
in the quantum condensed  regime.
(As a consequence neither energy nor momentum are conserved,
which is not unexpected as these results apply
to an \emph{open} quantum subsystem;
see \S\ref{Sec:EoM-Cond}.)
Specifically, this says that for a highly occupied momentum state
the rate of change of momentum for most subsets
is much less than the classical prediction for a given applied force.
The greatest reduction occurs when $n_A = N_{\bf a}/2$,
in which case it is $ {\cal O}( 2^{-N_{\bf a}})$.
Typically for low lying momentum states $N_{\bf a} = {\cal O}(10^2)$,
and so this reduction is quite significant,
exponentially small in fact.
If the entire state attempts the transition
as a rigid body with $n_A = N_{\bf a}$,
then the combinatorial coefficient is unity
and there is no reduction from the classical law.
If an individual boson attempts the transition, $n_A=1$,
then its rate of change of momentum is reduced
by a factor of $N_{\bf a}^{-1}$ from the value
given by Newton's second law of motion.

In summary, for a given shared non-local force and a given subset,
transitions from a momentum state are more frequent
if the subset is the entire momentum state
than if it consists of a fraction of the bosons in the state.
But there are $(N_{\bf a}-1)(N_{\bf a}-1)!$ ways of choosing subsets
that are not the entire state.
Hence the vast majority of uniformly chosen subsets result
in an exponentially reduced rate of change of momentum.
But of course this all depends on the size distribution of the subsets,
and this is derived in the following  \S\ref{Sec:gna}.

Since superfluid flow consists selectively of bosons
in highly occupied states,
this says that the rate of change of their momentum
is exponentially reduced.
It follows that there is a direct connection between
the conservation of entropy on an adiabatic trajectory,
the reduction in the rate of change of momentum
in the condensed regime,
and the loss of shear viscosity in superfluidity.

A second contribution to the reduction in the rate of change of momentum
of a subset
is the shared non-local force of the subset,
$n_A {\bf F}_{A} =  \sum_{j\in A} {\bf f}_j$.
This shared non-local  force is a consequence of the permutations
of the bosons in the momentum state.
The magnitude of the change due to this for the momentum state
scales with $n_A^{1/2}$,
whereas the dissipative effects on the classical shear viscosity
of individual momentum changes scale with $n_A$.
This effect is largest for the largest subsets,
but is non-existent for individual transitions (see next).
This non-local sharing effect
reduces the rate of change of momentum in shear flow
from its classical value.
The combinatoral effect of the occupation entropy
and the effect of the non-local sharing of the forces
explain at the molecular level
the reduction in viscosity in the condensed superfluid.

\subsection{Distribution of Subset Size}
\label{Sec:gna}

Let $\wp(n_A|N_{\bf a})$
be the probability of a subset of size $n_A$
irrespective of its composition
being involved in a transition
from the momentum state containing $N_{\bf a}$ bosons.
We now incorporate this into the unconditional transition probability.
As in the preceding section,
microscopic reversibility yields
\begin{eqnarray}
\lefteqn{
\wp({\bf \Gamma}'|{\bf \Gamma};n_A,\tau) \wp({\bf \Gamma})
\wp(n_A|N_{\bf a})
} \nonumber \\
& = &
\wp({\bf \Gamma}^\dag|{\bf \Gamma}'^\dag;n_A,\tau)
\wp({\bf \Gamma}'^\dag) \wp(n_A|N_{{\bf a}_\alpha'}') .
\end{eqnarray}
In the conjugate system, $N_{-{\bf a}}= N_{{\bf a}}$.
We write $N_{\bf a}' = N_{\bf a} - n_A$
and $N_{{\bf a}_\alpha'}' = N_{{\bf a}_\alpha'} + n_A$.

The current conditional transition probability given in \S\ref{Sec:AdTrans}
remains valid if the size probability cancels here,
\begin{equation}
\wp(n_A|N_{\bf a}) = \wp(n_A|N_{{\bf a}_\alpha'}').
\end{equation}
In general $ N_{\bf a} \ne N_{{\bf a}_\alpha'}'$,
and in view of the fact that the respective normalization constants
in general depend upon each occupancy,
and also the facts that $ N_{\bf a} \ge 1$
and $ N_{{\bf a}_\alpha'}'\ge 1$,
the only way to ensure equality here
is to choose
\begin{equation}
\wp(n_A|N_{\bf a}) = \delta_{n_A,1}.
\end{equation}
This is equivalent to choosing $g(n_A) = \delta_{n_A,1}$
in Eq.~(\ref{Eq:wploop}),
so that the conditional transition probability
for boson $j \in {\bf a}$  becomes
\begin{eqnarray} \label{Eq:wploop1}
\lefteqn{
\wp_j({\bf a}'_{j\alpha}|{\bf a},N_{\bf a})
}  \\ \nonumber
& = &
\frac{|\tau f_{j,\alpha}|}{N_{{\bf a}}\Delta_p}
\left\{ 1
- \frac{ \beta \Delta_p}{2m} \mbox{sign}(\tau {f}_{j\alpha}) a_\alpha
+ \frac{ \beta \tau a_\alpha }{2m}  {f}_{j\alpha}  \right\} .
\end{eqnarray}
Here  ${\bf a}'_{j\alpha} = {\bf a}
 + \mbox{sign}(\tau f_{j\alpha}) \Delta_p \widehat{\bf x}_\alpha$.
Sequential transitions for each component and for each
boson in the momentum state are attempted at each time step.
There appears to be little difference
in whether the occupancy is updated after each successful transition,
or only at the end of the time step.
There also appears little difference if the quadratic term in the change in
kinetic energy is added.
And it is also possible to use the exponential form of the term in braces.

The restriction to individual transitions,
$g(n_A) = \delta_{n_A,1}$,
means that the damping of the rate of change of momentum
is polynomial rather than exponential, $\propto N_{\bf a}^{-1}$.
Since the occupancy of low-lying momentum states
can by ${\cal O}(10^2)$ below the condensation transition,
this is still a significant reduction,
and it appears sufficient to be interpreted as
the origin of superfluidity.

\comment{ 

\ldots

Using the algorithm described in the text,
including ${\bf F}_A = n_A^{-1} \sum_{j\in A} {\bf f}_j$,
for $N=1,000$, $T^*=0.60$, and $\rho^*=0.8872$,
the results in the text are $N_{000}=140(14)$
compared to $N_{000}^\mathrm{id}=86.97$.
The respective occupancies of the first excited momentum state are
$N_{001}=16.3(10)$ and $N_{001}^\mathrm{id}=14.53$.
The pressure is $\beta p/\rho = 3.63(12)$
and the viscosity is $\eta^*(6) = 25.4(74)$.

For the same case but using individual transitions
$g(n_A) = \delta_{n_A,1}$ and the shared nonlocal force
${\bf F}_A = {\bf F}_{\bf a}$,
the results are $N_{000}=114(10)$, $N_{001}=16.0(9)$,
 $\beta p/\rho = 5.87(9)$, $\eta^*(6) = 27.3(70)$.

For the same case but using individual transitions
$g(n_A) = \delta_{n_A,1}$ and the local force
${\bf F}_A = {\bf f}_j$,
the results are $N_{000}=90.7(78)$, $N_{001}=15.8(7)$,
 $\beta p/\rho = 3.00(11)$, $\eta^*(6) = 15.6(48)$.

} 

\subsection{Dissipative Transition}

The dissipative transitions act like a thermostat
and provide another  mechanism
for the change in occupancy of the momentum states
and for the equilibration of the occupancy distribution.
For this I randomly or sequentially choose a boson
from the $N$ bosons in the subsystem, say $j$.
I use the following conditional transition probability
for the 27 near neighbor states ${\bf a}'$
(including the original state ${\bf a}$).
The block of $N$ individual attempted transitions
is performed typically once every 10 time steps,
although less frequent attempts would probably suffice.

The dissipative transition is irreversible,
which means that the forward and backward unconditional transitions
are equally likely.
Hence  for the transition to a neighboring momentum state
${\bf a} \stackrel{j}{\to} {\bf a}'$,
the ratio of conditional transition probabilities is
\begin{eqnarray}
\frac{\wp_j({\bf a}'|{\bf a})}{\wp_j({\bf a}|{\bf a}')}
& = &
\frac{\wp_j({\bf a}')}{\wp_j({\bf a})}
\nonumber \\ & = &
\frac{N_{{\bf a}'}+1}{N_{\bf a}}
\left[ 1  - \frac{\beta (a'^2-a^2)}{2m} \right] .
\end{eqnarray}
Here the change in kinetic energy has been expanded to quadratic order.
This is satisfied by
\begin{equation} \label{Eq:DissQuad}
\wp_j({\bf a}'|{\bf a})
=
\left\{ \begin{array}{ll}
\displaystyle
\frac{\varepsilon}{N_{\bf a}}
\left[ 1  - \frac{\beta (a'^2-a^2)}{4m} \right],
& {\bf a}' \ne {\bf a}  \\
\displaystyle
1 - \frac{26 \varepsilon}{N_{\bf a}}
+  \frac{54\beta \Delta_p^2}{4m} \frac{\varepsilon}{N_{\bf a}} ,
& {\bf a}' = {\bf a}  .
\end{array} \right.
\end{equation}
For the following results, $\varepsilon=1/27$.
The dissipative transitions were attempted one boson at a time,
for all $N$ bosons in a cycle.
A cycle of such attempts was made
once every {\tt skipcon} adiabatic time steps.
These  dissipative transitions 
roughly correspond to $r=1$,
$g(n_A) = \delta_{n_A,1}$,
and $\lambda=1/{\tt skipcon}$.

The present algorithm
has proven adequate to ensure the equilibrium distribution,
although it is not actually clear that a dissipative thermostat is required
because unlike the classical adiabatic equations of motion,
temperature already appears
in the present adiabatic conditional transition probability.

\subsubsection{Ideal Boson Results} \label{Sec:Res2}

For ideal bosons,
this second order dissipative transition algorithm
for $T=0.60$, $\rho=0.8872$,  and $N=1,000$,
gave for the occupancies of the first several low lying momentum states:
$ N_{000} = 85.9(43)$, (exact is 87.0),
$ N_{001} = 16.1(3)$, (exact is 14.5),
$ N_{011} = 8.2(1)$, (exact is 7.73),
$ N_{111} = 5.39(4)$, (exact is 5.17),
$ N_{002} = 3.94(3)$, (exact is 3.83).

For $T=0.60$, $\rho=0.8872$, and  $N=10,000$,
the results were
$ N_{000} = 138.2(284)$, (exact is 138.0),
$ N_{001} = 52.0(31) $, (exact is 51.9),
$ N_{011} = 33.3(8) $, (exact is 31.8),
$ N_{111} = 23.5(5) $, (exact is 22.8),
$ N_{002} = 18.3(6) $, (exact is 17.8).
These improve significantly
the results of the linear algorithm given in
earlier versions of this paper.

\subsection{Viscosity Time Correlation Function} \label{Sec:Visco}

The shear viscosity can be expressed as an integral of
the momentum-moment time-correlation function
(Attard 2012a Eq.~(9.117)),
\begin{equation} \label{Eq:eta(t)}
\eta_{\alpha\gamma}(t)
=
\frac{1}{2 V  k_\mathrm{B} T}
\int_{-t}^{t} \mathrm{d}t'\,
\left< \dot P_{\alpha\gamma}^0({\bm \Gamma})
\dot P_{\alpha\gamma}^0({\bm \Gamma}(t'|{\bm \Gamma},0))
\right> .
\end{equation}
It was Onsager (1931) who originally gave
the relationship between the transport coefficients
and the time correlation functions.
It is therefore somewhat puzzling
that this is called a Green-Kubo expression (Green 1954, Kubo 1966).
The first $\alpha$-moment of the $\gamma$-component of momentum is
\begin{equation}
P_{\alpha\gamma} = \sum_{j=1}^N q_{j\alpha} p_{j\gamma}.
\end{equation}
The classical adiabatic rate of change of momentum moment is
\begin{equation}
\dot P^0_{\alpha\gamma}
=
\frac{1}{m} \sum_{j=1}^N p_{j\alpha} p_{j\gamma}
+
\sum_{j=1}^N q_{j\alpha} f_{j\gamma} .
\end{equation}
The force ${\bf f}_{j}$
gives the classical rate of change of momentum of boson $j$.
This can be cast in symmetric form using the gradient of the pair potential,
in which form the minimum image convention can be applied.

In the condensed regime,
the average rate of change of momentum
for boson $j$ to leading order is
\begin{equation}
\left\langle \dot {\bf p}_{j}^0 \right\rangle
=
\frac{1}{N_{{\bf p}_j}} {\bf f}_j  .
\end{equation}
This uses $g(n_A) = \delta_{n_A,1}$,
Eq.~(\ref{Eq:wploop1}).
With this
the adiabatic rate of change of the first momentum moment is
\begin{eqnarray} 
\dot{\underline {\underline P}}^0
& = &
\frac{1}{m} \sum_{j=1}^N {\bf p}_{j} {\bf p}_{j}
+
\sum_{j=1}^N
{\bf q}_{j} \frac{1}{N_{{\bf p}_j}} {\bf f}_j
\nonumber \\ & = &
\frac{1}{m} \sum_{j=1}^N {\bf p}_{j} {\bf p}_{j}
+
\frac{1}{2} \sum_{j,k} \tilde {\bf q}_{jk} {\bf f}_{j,k} .
\end{eqnarray}
Here ${\bf f}_{j,k}$ is the force on boson $j$ due to boson $k$,
so that the total force on boson $j$ is
${\bf f}_{j} = \sum_k {\bf f}_{j,k}$,
and
\begin{equation}
\tilde {\bf q}_{jk}
\equiv
\frac{1 }{N_{{\bf p}_j}} {\bf q}_j
-
\frac{1 }{N_{{\bf p}_k}} {\bf q}_k .
\end{equation}

Generally simulations are conducted with periodic boundary conditions,
which give a homogeneous system.
In the uncondensed regime, $N_{\bf a} = 1$
and $\tilde {\bf q}_{jk}=  {\bf q}_{jk}$.
The minimum image separation is
${\bf q}_{jk} = {\bf q}_{jk} - L*{\tt ANINT}({\bf q}_{jk}/L)$,
where {\tt ANINT} gives the nearest integer.
This guarantees that $|q_{jk,\alpha}| \le L/2$
and it is known to work in the classical case
(ie.\ it gives a viscosity that is independent of system size).
In the condensed regime, $N_{{\bf p}_j} \ge 1$,
and so the minimum image convention can still be applied to
$\tilde {\bf q}_{jk}$ with the same guarantee.

\comment{ 
for  condensed bosons,
$n_A \ge 2$ and $n_B \ge 2$,
then in most cases $|\tilde {Q}_{AB,\alpha}|
\le |\tilde {Q}_{A,\alpha}| + |\tilde {Q}_{B,\alpha}|
\le L/4 + L/4$.
Since the weighted separation is always less than $L/2$
the minimum image convention makes no difference.
(In the exceptional case of $n_A=N_{\bf a}$,
the minimum image validly applies to the center of mass
of the momentum state.)
If however there is one condensed boson $A$ and one uncondensed boson $B$,
then $|\tilde {Q}_{B,\alpha}| \le L/2$.
If the minimum image convention is applied
to $\tilde {\bf Q}_{AB}$, it is the same as choosing
for $k$, the boson in $B$, from the images
$ \{ q_{k\alpha}, q_{k\alpha}+L, q_{k\alpha}-L \}$
to minimize the distance, which is legitimate.
The conclusion is that in all instances
the minimum image convention,
$\tilde {\bf Q}_{AB}
= \tilde {\bf Q}_{AB}
- L*{\tt ANINT}(\tilde {\bf Q}_{AB}/L)$,
gives correct results.
} 

In the simulations,
the adiabatic rate of change of the first momentum moment
was calculated once every 75 time steps.
The time correlation function was constructed and integrated
to give the viscosity time function.
This was averaged over the six components
(three in the classical case)
and the maximum value taken to be `the' shear viscosity.

%
\section{Results} \label{Sec:Res1}
\setcounter{equation}{0} \setcounter{subsubsection}{0}
%

In what follows what is defined as the classical fluid
is treated with the same algorithm as the quantum fluid
but as if the momentum states were solely occupied.
Hence individual transitions were attempted
without the factor of $N_{\bf a}^{-1}$ in the conditional probability
or in the rate of change of the first momentum moment in the shear viscosity.
However, the momenta were still quantized
and the adiabatic transitions were still stochastic,
just as in the quantum case.

The Lennard-Jones pair potential was used,
\begin{equation}
u(r) =  4 \varepsilon
\left[  \frac{\sigma^{12}}{r^{12}} - \frac{\sigma^{6}}{r^{6}} \right] .
\end{equation}
This was set to zero for $r> 3.5\sigma$.
The molecular diameter for helium is
$\sigma_\mathrm{He} = 0.2556$\,nm
and the well-depth is
$ \varepsilon_\mathrm{He}/k_\mathrm{B} = 10.22 $\,J
(van Sciver 2012).
The mass is $m_\mathrm{He} = 4.003\times 1.66054\times 10^{-27}$\,kg.

The results below are presented in dimensionless form:
the temperature is $T^* = k_\mathrm{B}T/\varepsilon_\mathrm{He}$,
the number density is $\rho^* = \rho \sigma_\mathrm{He}^3$,
the time step is $\tau^* = \tau /t_\mathrm{He}$,
the unit of time is
$t_\mathrm{He} =
\sqrt{m_\mathrm{He}\sigma_\mathrm{He}^2/\varepsilon_\mathrm{He}}$,
and the shear viscosity is
$\eta^* = \eta \sigma_\mathrm{He}^3/\varepsilon_\mathrm{He} t_\mathrm{He}$.

A canonical system, most commonly with $N=1,000$
Lennard-Jones atoms, was simulated with the above
quantum molecular dynamics algorithm.
The time step was $\tau^* = 2 \times 10^{-5}$.
A block of $N$ dissipative transitions
was attempted once every 10 time steps.
For the classical liquid
this combination consistently gave a kinetic energy
5-7\% lower than the exact value.
The present parameters  were judged acceptable
even though a smaller time step or a larger system
would have given more accurate results
albeit at the cost of longer simulation times.
More frequent dissipative transitions would also likely
improve the classical kinetic temperature
but at the cost of greater interference
in the adiabatic evolution of the momentum moment.
Averages were collected once every 75 time steps.
The number of time steps in a run was $10\times4,000\times75$.
A run took 24--30~hours on a desk-top personal computer.
Typically 4--8  runs were made at each thermodynamic point,
although twelve were made in the classical cases
at the lowest temperatures.
The quoted statistical error is the larger
of the error estimated from the fluctuations in 10 blocks of each run
(averaged across the runs),
or the standard deviation across the runs.
There was usually little difference in the two estimates (but see below).
The shear viscosity time function was estimated for $t^* \le 6$,
which meant that the time correlation function was covered ten times,
and that in a single run
$10 \times 6 $ values were averaged at each time point for the quantum case,
and $10 \times 3 $ values in the classical case.
In the quantum case the viscosity time function
had more or less reached it maximum by $t=6$.
In the classical case the viscosity time function was extrapolated beyond
$t=6$ to its maximum value using a quadratic best fit.
The quantum viscosity time function
was rather flat and no extrapolation was made.
The components of the quantized momentum were restricted to
$p_{j\alpha}^2/2mk_\mathrm{B}T \alt 20$.

The liquid saturation curve obtained in previous work
was followed (Attard 2023a):
$\{T^*,\rho^*\}=$
$\{1.00,0.7009\}$, $\{0.90,0.7503\}$, $\{0.80,0.8023\}$,
$\{0.75,0.8288\}$, $\{0.70,0.8470\}$, $\{0.65,0.8678\}$
and $\{0.60,0.8872\}$.

\begin{figure}[t]
\centerline{ \resizebox{8cm}{!}{ \includegraphics*{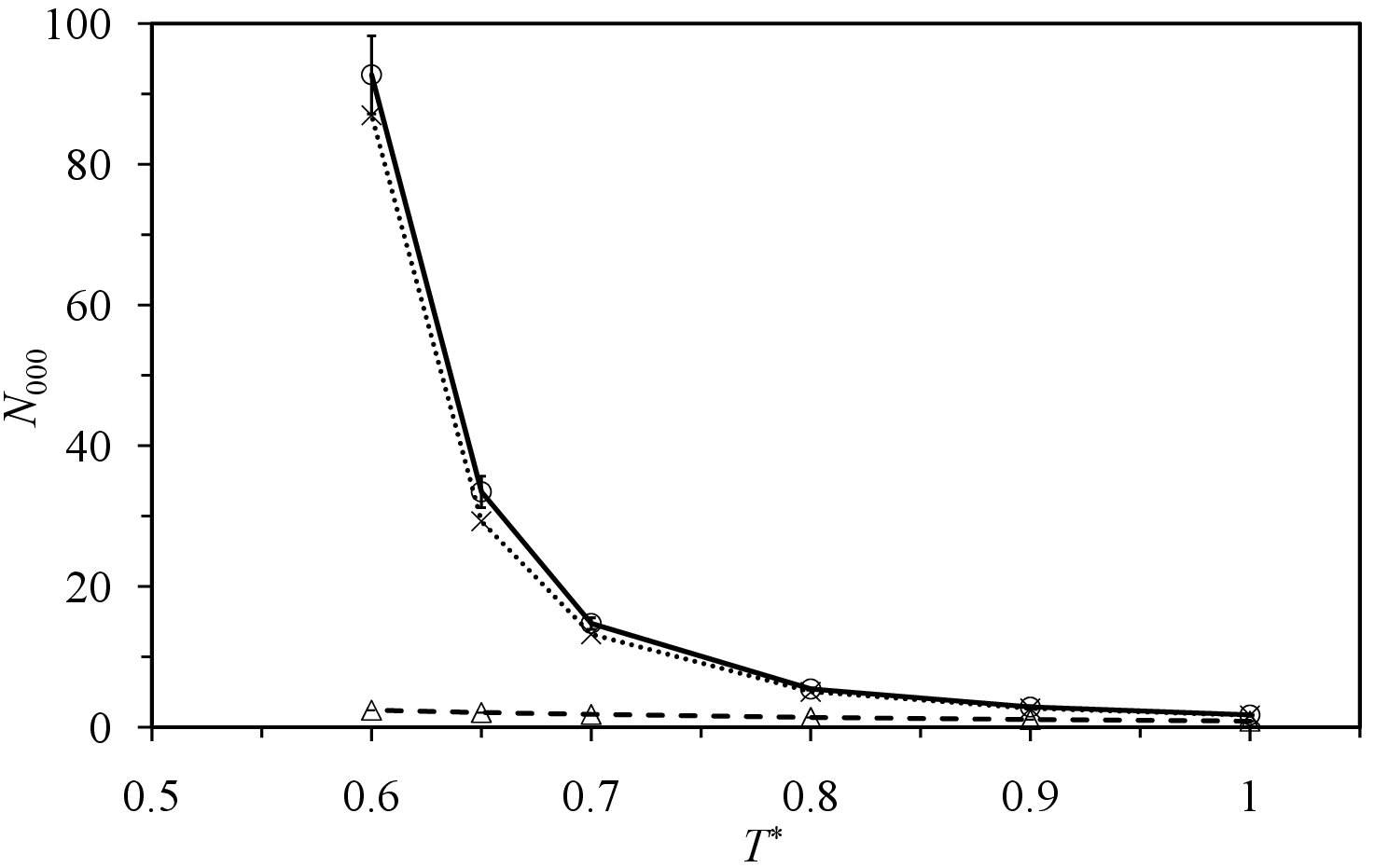} } }
\caption{\label{Fig:N000}
Ground state occupancy in the saturated Lennard-Jones liquid.
The circles are the quantum liquid,
the triangles are the classical liquid,
and the crosses are the exact result for ideal bosons.
The error bars are less than the symbol size.
The lines are an eye guide.
Note that $T[K] = 10.22T^*$.
}
\end{figure}

Figure~\ref{Fig:N000} shows the ground state occupancy
on the saturation curve for the Lennard-Jones liquid.
It can be seen that the simulated occupancy in the quantum liquid
is in good agreement with the analytic result calculated
for non-interacting bosons.
Comparable if not better agreement holds for the first several
excited momentum states.
The ideal boson result should apply to interacting bosons
on the far side of the $\lambda$-transition
(Attard 2025 \S5.3).
The slightly larger than ideal value
for the quantum liquid ground momentum state occupancy
is probably a finite size effect.

At lower temperatures it can be seen that
the occupancy  in the quantum liquid is much larger
than for the classical liquid.
This is of course due to the role of the occupation entropy
on the transition probability.
However, the occupancy of the ground state in the quantum liquid
is a small fraction of the total number of bosons in the subsystem.
These fractions decrease with increasing subsystem size.
Obviously this means that ground state condensation
cannot account for the $\lambda$-transition
or for superfluidity.

\begin{figure}[t]
\centerline{ \resizebox{8cm}{!}{ \includegraphics*{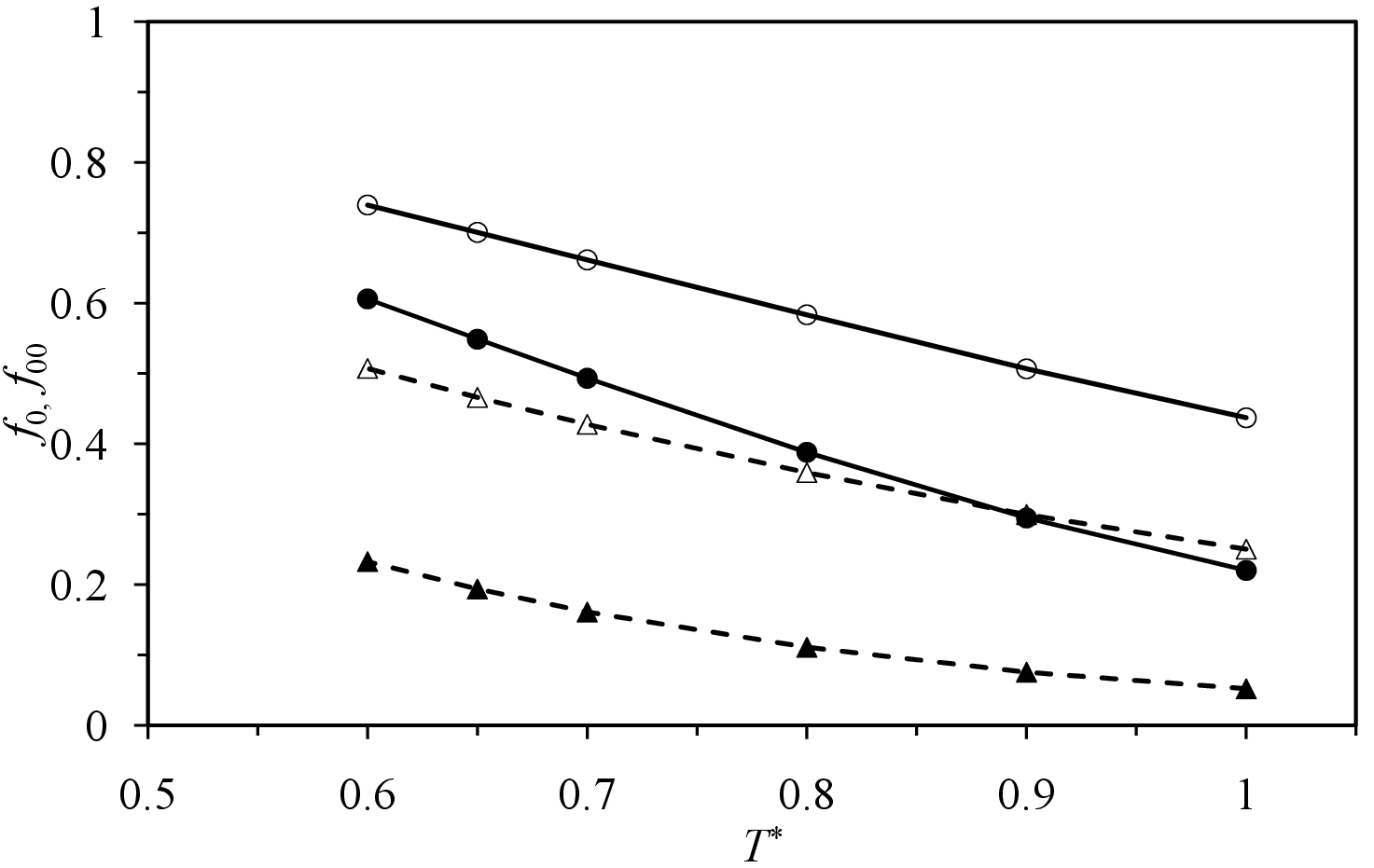} } }
\caption{\label{Fig:f0}
Fraction of bosons condensed in the saturated Lennard-Jones liquid.
The circles are the quantum liquid,
the triangles are the classical liquid.
The open symbols are $f_0$,
the fraction in states occupied by two or more bosons,
and the filled symbols are $f_{00}$,
the fraction in states occupied by three or more bosons.
The error bars are less than the symbol size.
The lines are an eye guide.
}
\end{figure}

Figure~\ref{Fig:f0} gives the fraction of condensed bosons
in the subsystem.
Condensed bosons were defined as those in multiply occupied states,
with a threshold set at 2 for $f_0$ and 3 for $f_{00}$.
In the quantum liquid at the lowest temperature studied
about 74\% of the bosons are in states with two or more,
and about 61\% are in states with three or more.
In contrast the fraction for the classical liquid
is 51\% and 23\%, respectively.
These specific results are for $N=1,000$,
but other simulations show that these fractions are quite insensitive
to the system size.
The conclusion is that at these temperatures
Bose-Einstein condensation is substantial,
and that multiple momentum states are multiply occupied.
That the majority of the bosons in the system can be considered
to be condensed explains the macroscopic nature
of the $\lambda$-transition and superfluidity.

It can be seen that at higher temperatures
the condensation in the quantum liquid
is approaching that in the classical liquid,.
However even at the highest temperature studied, $T^*=1.00$,
there is still excess condensation in the quantum liquid,
$f_0^\mathrm{qu}=44\%$ compared to $f_0^\mathrm{cl}=25\%$.
That there remains condensation in the quantum liquid
well-above the superfluid transition temperature
is likely due to the neglect in the present calculations
of position permutation loops,
which suppress condensation (Attard 2025 \S3.2).

The kinetic energy per boson in the classical liquid is
$\beta {\cal K}/N=1.4513(4)$ at $T^*=1.00$
and 1.4048(2) at $T^*=0.60$.
The equipartition theorem gives the exact classical value as 3/2.
Clearly the present stochastic equations of motion
that use the transition probability for quantized momentum
are close to the continuum classical equations of motion.
The discrepancy is probably an effect of finite size.
The kinetic energy per boson in the quantum liquid is
$\beta {\cal K}/N=1.2248(5)$ at $T^*=1.00$
and 0.778(2)  at $T^*=0.60$.
The decrease in kinetic energy with decreasing temperature
is a manifestation of the increasing condensation
in the quantum liquid
that preferentially occurs in the low lying momentum states.

\begin{figure}[t]
\centerline{ \resizebox{8cm}{!}{ \includegraphics*{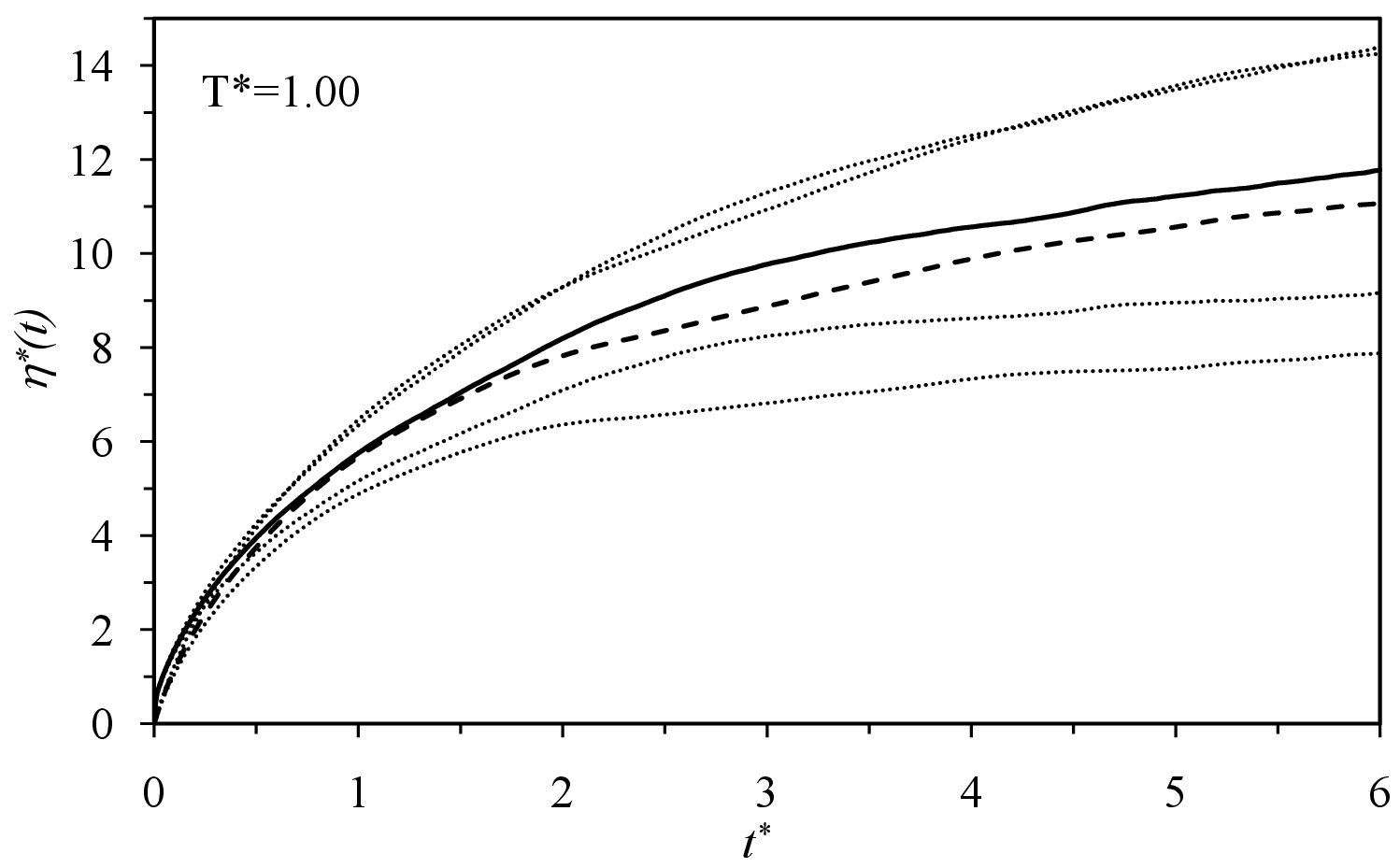} } }
\caption{\label{Fig:eta100}
Shear viscosity time function for the Lennard-Jones liquid
at $T^*=1.00$ and $\rho^*=0.7009$.
The solid curve is the quantum liquid,
the dashed curve is the classical liquid,
and the dotted curves give the 95\% confidence level.
}
\end{figure}

Figure~\ref{Fig:eta100} shows the shear viscosity time function
for the quantum and classical liquids at the relatively high temperature
of $T^*=1.00$.
The general features of the curves are that they rise from zero at $t^*=0$
to reach a maximum or a plateau, in this case at about $t^*=6$.
This maximum value is taken to be `the' shear viscosity.
In this and lower temperatures the quantum viscosity
is rather flat by $t^*=6$;
compare $\eta^\mathrm{qu}(6) = 11.8(26)$
with the extrapolated maximum $\eta^\mathrm{qu}(10.9) = 13.0$.
At this highest temperature the same applies to the classical viscosity;
compare $\eta^\mathrm{cl}(6) = 11.1(32)$
with the extrapolated maximum $\eta^\mathrm{cl}(7.3) = 11.3(34)$.

It can be seen that the viscosities of the quantum and classical liquids
are statistically indistinguishable at this highest temperature studied.
The fraction of condensed bosons in states with two or more bosons
at this temperature
is 44\% for the quantum liquid and 25\% for the classical liquid
(Fig.~\ref{Fig:f0}).
The average occupancy of momentum states containing two or more bosons
in the quantum liquid was
$\overline N^\mathrm{qu,0}_\mathrm{occ} = 2.6258(5)$,
and in the classical liquid it was
$\overline N^\mathrm{cl,0}_\mathrm{occ} = 2.1640(4)$.
(The average occupancy of occupied states was
$\overline N^\mathrm{qu}_\mathrm{occ} = 1.3712(3)$,
and $\overline N^\mathrm{cl}_\mathrm{occ} = 1.1557(1)$, respectively.)
Evidently at this temperature
the occupation entropy has little effect on the occupation of
momentum states.
And the reduction in the rate of change of momentum due to occupancy
has little effect on the viscosity.

The diffusion in the system was relatively small.
The root mean square change in separation
of a randomly chosen pair of bosons
over the course of a run was 0.83(23) for the quantum liquid
and 0.63(7) for the classical liquid.
The statistical errors in the average potential energy and pressure
were larger when estimated  between runs rather than within a run.
(Each run was started from an independently equilibrated configuration.)
This suggests that the subsystem is somewhat glassy.

\begin{figure}[t]
\centerline{ \resizebox{8cm}{!}{ \includegraphics*{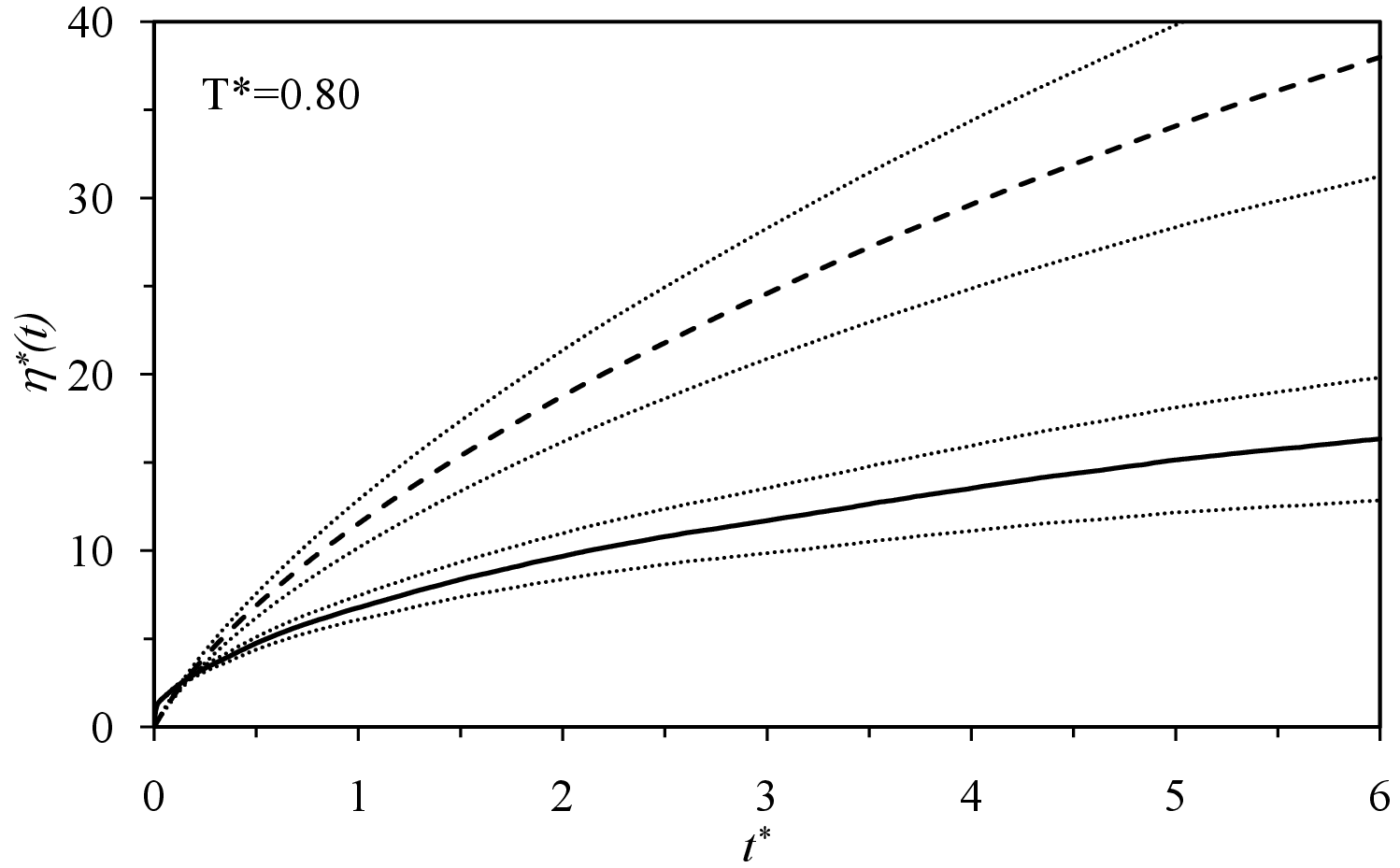} } }
\caption{\label{Fig:eta80}
As in the preceding figure but for $T^*=0.80$ and $\rho^*=0.8023$.
}
\end{figure}

Figure~\ref{Fig:eta80} gives the shear viscosity time function
at the lower temperature of $T^*=0.80$.
Note the change of scale.
The viscosity of the quantum liquid is significantly lower
than that of the classical liquid over virtually the whole domain.
At very small times the quantum viscosity rises more quickly
than the classical viscosity before flattening out.
This effect was seen consistently in all the simulations,
but the quantitative values in the small time regime appear sensitive
to the frequency of the application of the dissipative thermostat.
The extrapolated maximum viscosity of the classical liquid,
$\eta^\mathrm{cl}(12)=49(13)$
(compare $\eta^\mathrm{cl}(6)=38.0(68)$),
is much greater at this temperature
than that at the higher temperature of the preceding figure.
The root mean square change in separation was
0.45(8) for the quantum liquid and 0.59(9) for the classical liquid.
The statistical errors in the average potential energy and pressure
in the classical case
were less when estimated from within a run
as from between runs,
suggesting a relatively glassy subsystem.

\begin{figure}[t]
\centerline{ \resizebox{8cm}{!}{ \includegraphics*{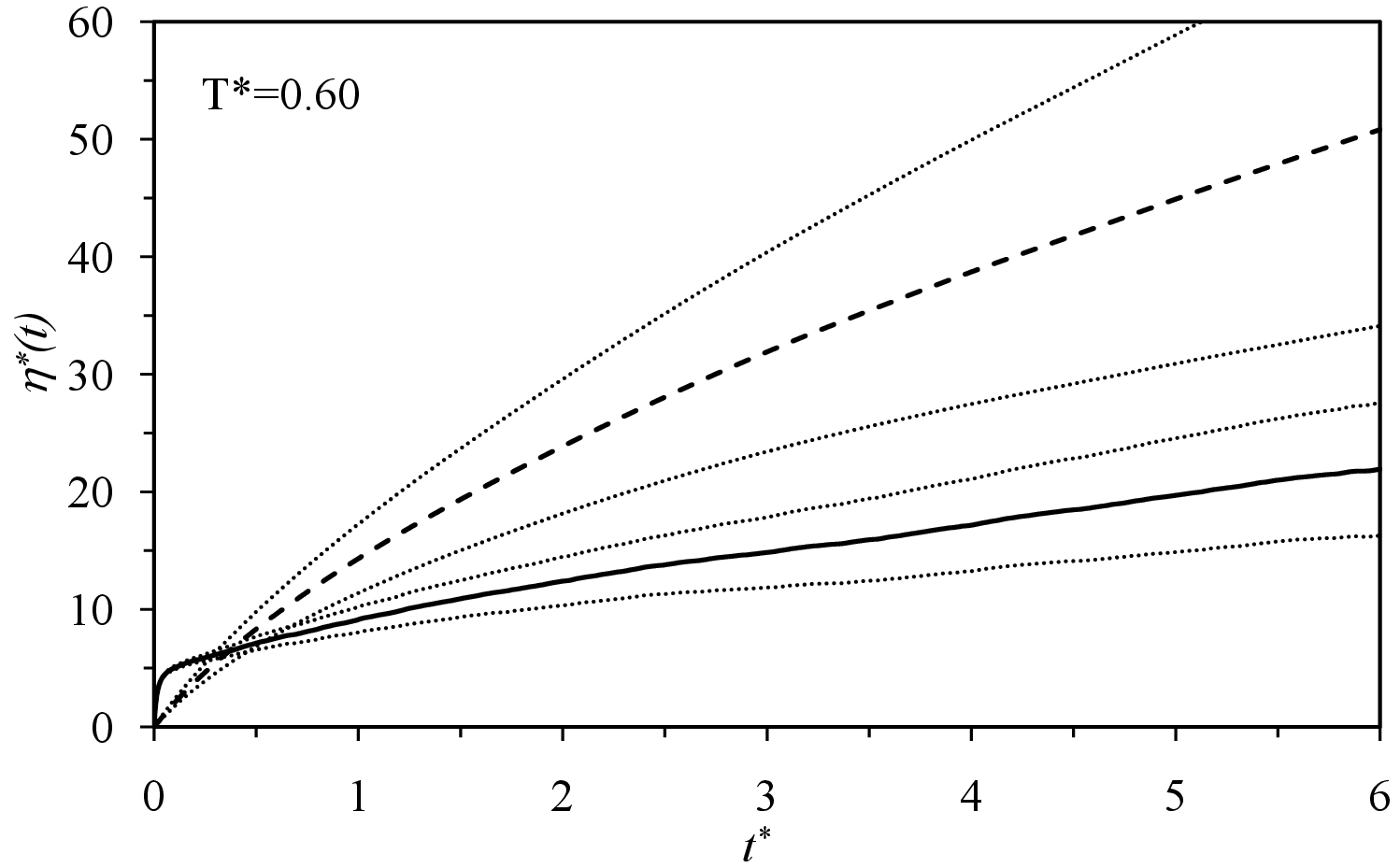} } }
\caption{\label{Fig:eta60}
As in the preceding figure but for $T^*=0.60$ and $\rho^*=0.8872$.
}
\end{figure}

Figure~\ref{Fig:eta60} shows the viscosity for the
lowest temperature studied,  $T^*=0.60$.
Note the change of scale again.
It can be seen that the classical viscosity is substantially larger
than the quantum viscosity everywhere except at very short times.
In the quantum case the viscosity time function is relatively flat
at the terminus, which suggests that it is close to its maximum,
$\eta^\mathrm{qu}(6)=21.9(57)$.
In the classical case,
there is enough curvature and slope to extrapolate to a maximum
$\eta^\mathrm{cl}(20)=91(53)$,
which is substantially larger than the value at the terminus,
$\eta^\mathrm{cl}(6)=51(17)$.

The root mean square change in separation over a run was
0.31(6) for the quantum liquid and 0.25(5) for the classical liquid.
The statistical errors in the average potential energy and pressure
were smaller when estimated within a run
than between runs.
Again the conclusion is that the subsystem is somewhat glassy.
If this is the case then it is remarkable that  the shear viscosity
has a finite value, and that it is so low for the quantum liquid.

\begin{figure}[t]
\centerline{ \resizebox{8cm}{!}{ \includegraphics*{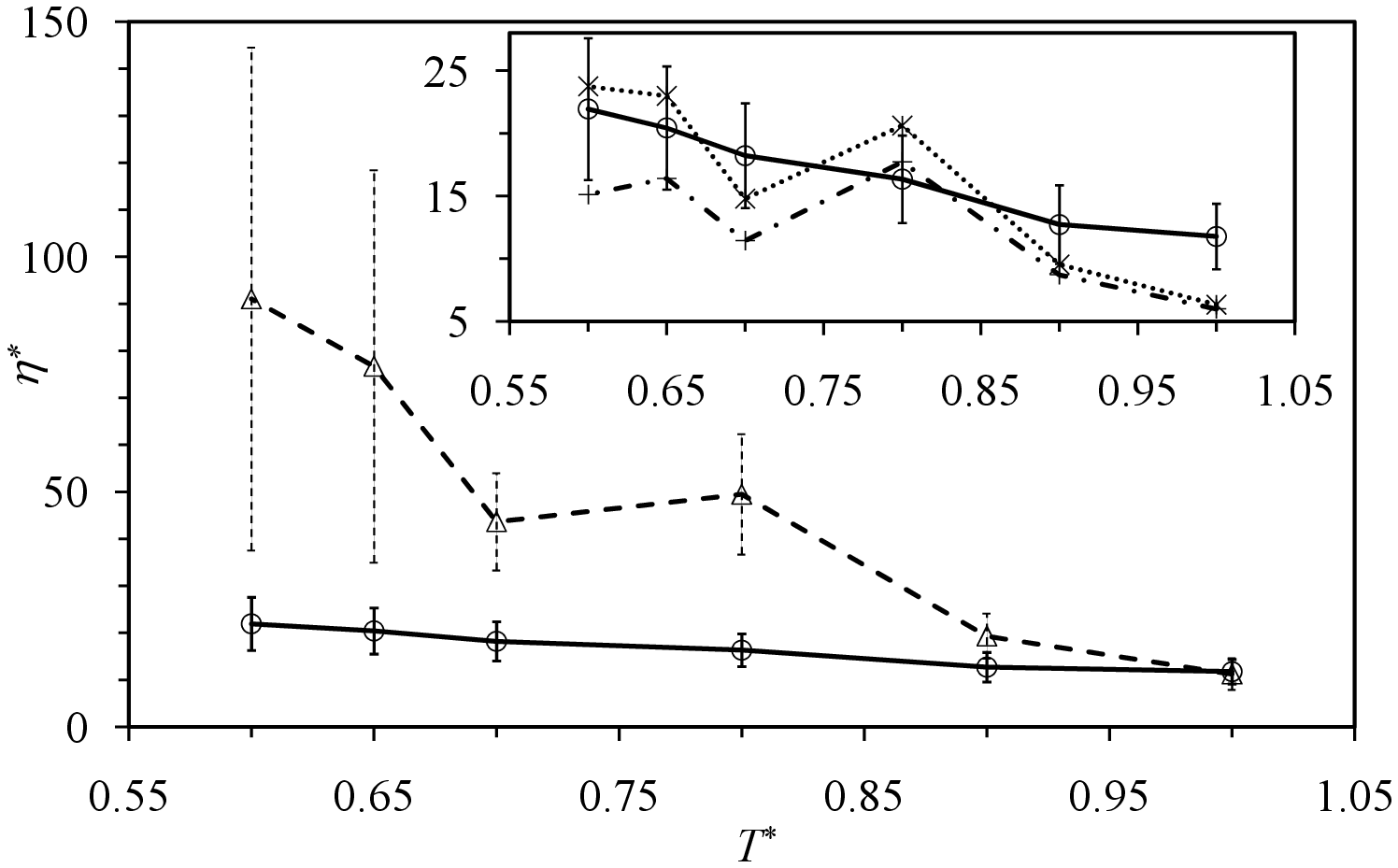} } }
\caption{\label{Fig:eta}
Shear viscosity on the Lennard-Jones liquid saturation curve.
The circles are for the quantum liquid
$\eta^\mathrm{qu}(6)$,
the triangles are for the classical liquid
$\eta^\mathrm{cl}_\mathrm{max}$.
The error bars give the 95\% confidence level,
and the lines are an eye guide.
\textbf{Inset.}
Viscosity of the quantum liquid (circles) on a magnified scale
compared to two predictions based on the classical viscosity.
The asterisks connected by dotted lines are  Eq.~(\ref{Eq:etaT}),
and the plus symbols connected by dash-dot lines
are Eq.~(\ref{Eq:etaA}).
The statistical error of these is larger
than that shown for the quantum liquid.
}
\end{figure}

Figure~\ref{Fig:eta} shows the shear viscosity
as a function of temperature for the saturated liquid.
Based on the flatness of the curves,
the quantum viscosity evaluated at $t=6$
is close to its maximum value.
The classical viscosity maximum
is based on a quadratic best fit extrapolation.
It can be seen that at higher temperatures
the classical and quantum viscosities converge.
At the lowest temperature studied the classical viscosity is about
four times larger than the quantum viscosity.
Whereas the classical viscosity increases by a factor of eight
over the temperature range,
the quantum viscosity only increases by a factor of two.
The not quite doubling in condensation in the quantum liquid,
Fig.~\ref{Fig:f0},
is sufficient to cancel almost completely
the classical viscosity increase.
Evidently condensation reduces the rate of change of momentum
via the factor of $1/N_{\bf a}$,
and this directly reduces the shear viscosity.

The inset to the figure includes results for
\begin{equation} \label{Eq:etaT}
\eta_\mathrm{T}
=
(1-f_0^\mathrm{qu})\eta^\mathrm{cl}_\mathrm{max} .
\end{equation}
This viscosity is the analogue
of the two-fluid model of superfluidity (Landau 1941, Tisza 1938),
namely it is the fraction of uncondensed bosons
(ie.\ those in singly-occupied momentum states)
in the quantum liquid
times the viscosity in the classical liquid
(ie.\ the viscosity calculated as if all the bosons
were in singly-occupied momentum states).
Essentially it assumes that the viscosity of condensed bosons is zero,
and it assumes that the actual viscosity is a linear combination
of that of the individual components of a two-component mixture.

It can be seen that the two-fluid approximation is surprisingly good.
For $T^*=0.60$, $\rho^*=0.8872$,
the quantum viscosity is $\eta^\mathrm{qu}(6)=21.9(57)$,
and the linear binary mixture result is
$\eta_\mathrm{T}=24(14)$.
For $T^*=1.00$, $\rho^*=0.7009$,
the quantum viscosity is $\eta^\mathrm{qu}(6)=11.8 (26)$,
and the linear binary mixture result is
$\eta_\mathrm{T}=6.3(19)$.

Of course in the present equations of motion
the rate of changed of momentum of condensed bosons is not zero,
but is rather reduced by the occupancy of their respective momentum states.
For the lowest temperature studied,
$T^*=0.60$ $\rho^*=0.8872$,
the average occupancy of momentum states
in the quantum case was $\overline N_\mathrm{occ}  = 2.455(4)$.
For such a small average occupancy it is perhaps surprising
that the reduction of the rate of change of momentum
is sufficient to reduce the superfluid viscosity by so much.
In fact however a plausible model for the viscosity
in the condensed regime is
\begin{eqnarray} \label{Eq:etaA}
\eta_\mathrm{A}
& = &
\frac{1}{ \overline N_\mathrm{occ}^{2} }
\eta^\mathrm{cl}_\mathrm{max}.
\end{eqnarray}
Here
$\overline N_\mathrm{occ} \equiv  N/\overline M_\mathrm{occ}$
is the average occupancy of occupied states,
where $\overline M_\mathrm{occ}$ is the number of occupied states.
This factor gives the reduction in the rate of change
of the first momentum moment in the force term.
This neglects the diffusive (ie.\ ideal) contribution.
This is squared because the viscosity time function
is the \emph{pair} time correlation
of the rate of change of the first momentum moment.
In the case $T^*=0.60$ $\rho^*=0.8872$ this formula gives
$\eta_\mathrm{A}  =15.1(89)$,
to be compared with the two-fluid model $\eta_\mathrm{T}  =23.7(139)$
and the actual simulated value $\eta^\mathrm{qu}(6)=21.9(57)$.
At $T^*=1.00$ $\rho^*=0.7009$,
with $\overline N_\mathrm{occ} = 1.3712(3)$,
the respective values are
$\eta_\mathrm{A} = 6.0(18)$, $\eta_\mathrm{T} = 6.3(19)$,
and $\eta^\mathrm{qu}(6) = 11.8(26)$.
It can be seen in the inset to Fig.~\ref{Fig:eta}
that the accuracy of this second model is comparable
to that of the two-fluid model.
This second model has a quantitative justification
that does not insist that the viscosity of condensed bosons is zero.
This formula explains how the seemingly small average occupancy
of condensed bosons is sufficient to cause the reduction
in superfluid viscosity comparable to the measured values.

Of course the virtue of  molecular dynamics simulations such as the present
is that they enable the classical and quantum viscosities,
as well as the distribution of occupancies,
to be computed at any thermodynamic state point,
and in molecular detail.
One wonders how useful the two-fluid model really is,
since it does not appear possible to actually measure
the individual viscosities or the individual mole fractions
in a laboratory.

\begin{figure}[t]
\centerline{ \resizebox{8cm}{!}{ \includegraphics*{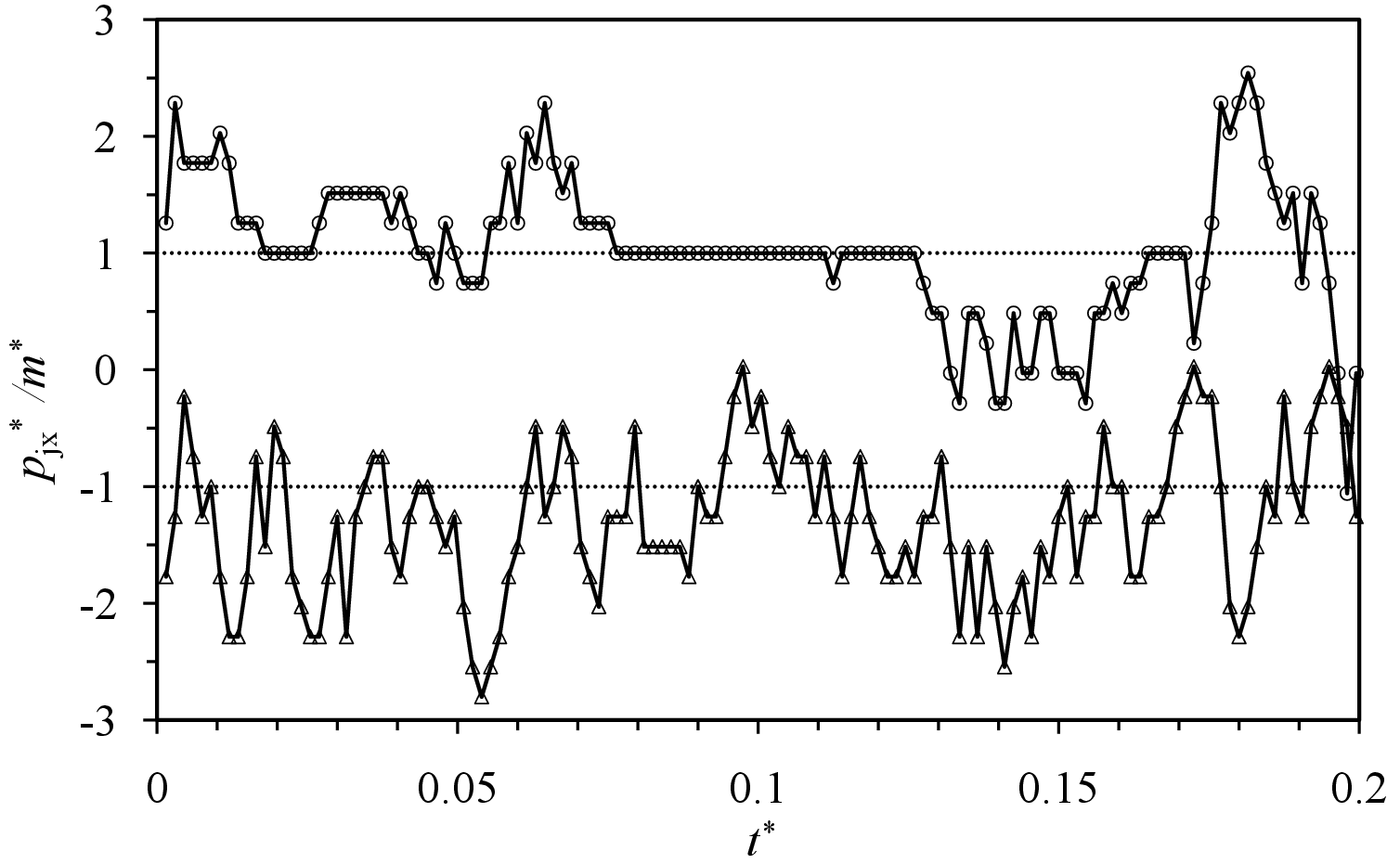} } }
\caption{\label{Fig:traj}
Component of velocity of a typical boson on a trajectory
at $T^*=0.60$ and $\rho^*=0.8872$, plotted once every 75 time steps.
The circles are for the quantum liquid
(offset by $+1$)
and the triangles are for the classical liquid
(offset by $-1$).
The dotted lines give the zero momentum state for the component.
The dimensionless spacing between momentum states is 0.26.
}
\end{figure}

Figure~\ref{Fig:traj}
shows a quantum and  classical trajectory
for a component of momentum of a typical boson.
It can be seen that there is a qualitative difference between
the trajectory in the quantum liquid and in the classical liquid.
The quantum equations of motion yield
a smoother curve, with smaller fluctuations,
and noticeable stretches of constant momentum.
These are correlated with the boson being
in a highly occupied momentum state,
which are noticeably the low-lying momentum states.
On this portion of the quantum trajectory,
the occupancy of the momentum state that this boson is in
ranges up to 91, and averages 19.4.
The conclusion is that the occupancy factor, $1/N_{\bf a}$,
damps the accelerations experienced by condensed bosons.
It is not hard to imagine that the more frequent changes in momentum
evident in the classical liquid dissipate momentum more efficiently
and give rise to the usual viscosity of everyday experience.

%
\section{Conclusion} \label{Sec:Concl2}
\setcounter{equation}{0} \setcounter{subsubsection}{0}
%

The two-fluid model of superfluidity (Landau 1941, Tisza 1938)
holds that below the $\lambda$-transition
helium is a mixture of helium I and helium II,
with the former having normal viscosity
and the latter being the superfluid with zero viscosity.
For a bulk measurement,
the shear viscosity is believed to be a linear combination of the two
in proportion to their mole fraction.
Helium II is believed to consist of bosons
condensed into the ground energy state.

My definition of condensation is different
namely that it occurs in the low-lying momentum states (Attard 2025).
I also say that it is a simplification to divide the liquid
into a mixture of two distinct fluids since
there are multiple multiply-occupied states.
Also, as few as two bosons
in the same momentum state experience the effects permutation symmetry.
Despite the  simplicity of the two-fluid model,
provided that condensation is interpreted
as being into the low-lying momentum states,
the results in Fig.~\ref{Fig:eta} provide a measure of support for it.

Elsewhere (Attard 2025 \S5.3)
I have argued that on the far side of the $\lambda$-transition
$^4$He is dominated by momentum loops,
which is to say permutations between bosons in the same momentum state.
In this case the quantum statistical partition function factorizes
into an ideal momentum part
and a classical position configuration integral.
This predicts that the occupancies of the momentum states
are given by ideal statistics,
for which the exact analytic results are known
(Attard 2025 \S2.2, Pathria 1972 \S7.1).
The present quantum stochastic molecular dynamics results
for the occupancy agree with ideal statistics.
This is a good test of the computational algorithm and programming.

The present analysis gives an explanation
for the molecular origin of superfluidity.
The shear viscosity is much reduced or practically zero
for condensed bosons because Newton's second law of motion does not hold:
the rate of change of momentum
due to an applied force is significantly smaller
in the condensed compared to the non-condensed regime.
These quantum molecular dynamics equations
are necessary for the equilibrium system to evolve at constant entropy,
as demanded by fundamental statistical considerations (Attard 2012a Ch.~7),
and as confirmed by the thermodynamic analysis
of fountain pressure measurements
of superfluid helium (Attard 2025 \S 4.6).
In addition,
for the case of a subset of bosons changing momentum state,
the applied force effective on each individual
boson is the shared non-local force,
which is reduced from the individual classical force
by averaging over the subset 
to which the boson belongs.

Although this paper has focussed on the formulation of
a computational algorithm for the viscosity of a quantum liquid,
it should not be assumed that the molecular equations of motion
that are given are merely a numerical technique.
The equations of motion yield real trajectories for particles,
and there is reason to suppose that
the transition probability reflects the underlying reality of nature.
This of course has implications
for the physical interpretation of quantum mechanics
and of the universe.
To which end I suggest
\begin{equation}
\mbox{speculate, calculate, and mensurate.}
\end{equation}

\section*{References}


\begin{list}{}{\itemindent=-0.5cm \parsep=.5mm \itemsep=.5mm}

\item 
Attard  P 2012a
\emph{Non-equilibrium thermodynamics and statistical mechanics:
Foundations and applications}
(Oxford: Oxford University Press)

\item 
Attard P 2018
Quantum statistical mechanics in classical phase space. Expressions for
the multi-particle density, the average energy, and the virial pressure
arXiv:1811.00730

\item 
Attard P 2021
\emph{Quantum Statistical Mechanics in Classical Phase Space}
(Bristol: IOP Publishing)

\item 
Attard P 2023a
\emph{Entropy beyond the Second Law:
Thermodynamics and statistical mechanics for equilibrium, non-equilibrium,
classical, and quantum systems}
(Bristol: IOP Publishing, 2nd edition)

\item 
Attard P 2023b
Quantum stochastic molecular dynamics simulations
of the viscosity of superfluid helium
arXiv:2306.07538

\item 
Attard P 2023d
Hamilton's equations of motion from Schr\"odinger's equation
arXiv:2309.03349

\item 
Attard P 2025
\emph{Understanding Bose-Einstein Condensation,
Superfluidity, and High Temperature Superconductivity}
(London: CRC Press)

\item 
Bohm D 1952
A suggested interpretation of the quantum theory
in terms of `hidden' variables. I and II.
\emph{Phys.\ Rev.}\ {\bf 85} 166 


\item 
de Broglie L 1928
La nouvelle dynamique des quanta,
in Solvay   p.~105

\item 
Caldeira A O and Leggett A J 1983
Quantum tunnelling in a dissipative system
\emph{Ann.\ Phys.}\ {\bf 149} 374

\item 
Goldstein S 2024
Bohmian mechanics
\emph{The Stanford Encyclopedia of Philosophy}
(Summer 2024 Edition), E N Zalta and U  Nodelman (eds.)
{\tt URL =
<https://plato.stanford.edu/archives/\\
sum2024/entries/qm-bohm/>}

\item 
Green M S 1954
Markoff random processes and the statistical mechanics
of time-dependent phenomena.
II. Irreversible processes in fluids.
\emph{J.\ Chem.\ Phys.}\ {\bf 23}, 298

\item 
Joos E and Zeh H D 1985
The emergence of classical properties through
interaction with the environment
\emph{Z.\ Phys.}\  B {\bf 59} 223


\item 
Kubo R 1966
The fluctuation-dissipation theorem
\emph{Rep.\ Prog.\ Phys.}\ {\bf 29} 255

\item 
Landau L D 1941
Theory of the superfluidity of helium II
\emph{Phys.\ Rev.}\ {\bf 60} 356



\item 
Mermin D 1989
What's wrong with this pillow
\emph{Physics Today} {\bf 42} 9

\item 
Mermin D 2004
Could Feynman have said this
\emph{Physics Today} {\bf 57} 10

\item 
Merzbacher E 1970
\emph{Quantum Mechanics} 2nd edn
(New York: Wiley)

\item 
Messiah A 1961
\emph{Quantum Mechanics}
(Amsterdam: North-Holland volumes 1 and 2)

\item 
Onsager L (1931)
Reciprocal relations in irreversible processes. I.
\emph{Phys.\ Rev.}\ {\bf 37} 405.
Reciprocal relations in irreversible processes. II.
\emph{Phys.\ Rev.}\ {\bf 38} 2265

\item 
Pathria R K 1972
\emph{Statistical Mechanics} (Oxford: Pergamon Press)

\item 
Schlosshauer M 2005
Decoherence, the measurement problem,
and interpretations of quantum mechanics
arXiv:quant-ph/0312059v4

\item 
van Sciver  S W 2012
\emph{Helium Cryogenics}
(New York: Springer 2nd edition)

\item  
Tisza L 1938
Transport phenomena in helium II
\emph{Nature} {\bf 141} 913



\item 
Zurek W H 1991
Decoherence and the transition from quantum to classical
\emph{Phys.\ Today} {\bf 44} 36

\item 
Zurek W H,  Cucchietti F M, and  Paz J P 2003
Gaussian decoherence from random spin environments
arXiv:quant-ph/0312207.



\end{list}



\appendix
%
\section{Occupation of energy states?}
\setcounter{equation}{0} \setcounter{subsubsection}{0}
\renewcommand{\theequation}{\Alph{section}.\arabic{equation}}
%

The text invoked the occupation of low-lying momentum states.
This contrasts with the conventional understanding
of Bose-Einstein condensation,
which maintains that condensation occurs into the ground energy state.
This appendix proves that for interacting bosons
the occupation of energy states is undefined.

Consider a system of $N$ interacting bosons,
which, for simplicity, has just two energy states
with energy per boson $e_1$ and $e_2$.
If the occupancy picture is valid then the energy eigenvalues
are of the form $E=N_1e_1 + N_2 e_2$,
where $N_a$ is the number of bosons in state $a$.
The energies per boson, $e_1$ and $e_2$, must be independent of occupancy.

If I add an extra boson and place it in the first state,
$N \Rightarrow N+1$, $N_1 \Rightarrow N_1+1$,
then the second state must be unaffected.
This implies that the eigenfunction of the second state
must be independent of the new particle and its position.
Hence the energy eigenfunction must factorize into those of the two states,
\begin{equation}
\phi_{N_1,N_2}({\bf q}^{N_1},{\bf q}^{N_2})
=
\phi_{1;N_1}({\bf q}^{N_1})
\phi_{2;N_2}({\bf q}^{N_2})  .
\end{equation}
This implies the energy eigenvalue equations
\begin{eqnarray}
\hat{\cal H}_{1;N_1}({\bf q}^{N_1}) \phi_{1;N_1}({\bf q}^{N_1})
& = &
N_1 e_1 \phi_{1;N_1}({\bf q}^{N_1})
\nonumber \\
\hat{\cal H}_{2;N_2}({\bf q}^{N_2}) \phi_{2;N_2}({\bf q}^{N_2})
& = &
N_2 e_2 \phi_{2;N_2}({\bf q}^{N_2}) .
\end{eqnarray}
But these require the total Hamiltonian operator
to be the sum of the individual Hamiltonian operators
\begin{equation}
\hat{\cal H}({\bf q}^{N})
=
\hat{\cal H}_{1;N_1}({\bf q}^{N_1})
+
\hat{\cal H}_{2;N_2}({\bf q}^{N_2})   .
\end{equation}
The interaction potential energy precludes this,
$U({\bf q}^{N}) \ne U({\bf q}^{N_1}) + U({\bf q}^{N_2})$.


There is an alternative way to see
why factorization is forbidden by interaction.
Using an abbreviated notation
and assuming a factorized eigenfunction
the energy eigenvalue equation is
\begin{equation}
\hat {\cal H} ({\bf q}_1,{\bf q}_2) \phi_1({\bf q}_1) \phi_2({\bf q}_2)
= [E_1+E_2] \phi_1({\bf q}_1) \phi_2({\bf q}_2) ,
\end{equation}
or
\begin{equation}
\frac{\hat {\cal K} ({\bf q}_1) \phi_1({\bf q}_1)}{\phi_1({\bf q}_1)}
+
\frac{\hat {\cal K} ({\bf q}_2) \phi_2({\bf q}_2)}{\phi_2({\bf q}_2)}
+ U({\bf q}_1,{\bf q}_2)
= E_1+E_2  .
\end{equation}
This follows since the kinetic energy operator
is the sum of single-particle operators.
Adding and subtracting this equation
for states ${\bf q}_1'$ and ${\bf q}_2'$ gives
\begin{equation}
U({\bf q}_1,{\bf q}_2) - U({\bf q}_1',{\bf q}_2)
- U({\bf q}_1,{\bf q}_2') + U({\bf q}_1',{\bf q}_2')
= 0 .
\end{equation}
But for nearby states this is just
the second cross derivative of the potential,
$({\bf q}_2'-{\bf q}_2)({\bf q}_1'-{\bf q}_1)
: \nabla_1 \nabla_2 U({\bf q}_1,{\bf q}_2)$.
By definition this is non-zero for interacting particles.
This proves that a factorized energy eigenfunction
can only hold for non-interacting particles.

I conclude that it is meaningless to speak of the occupation
of energy states for interacting particles.

%
\section{Momentum eigenfunctions}
\setcounter{equation}{0} \setcounter{subsubsection}{0}
%

In the text, and indeed in all of my work
on quantum statistical mechanics from the beginning (Attard 2018, 2021),
I take the  momentum eigenfunctions to be
(normalized, unsymmetrized)
\begin{equation}
\phi_{\bf p}({\bf q}) =
\frac{1}{V^{N/2}} e^{-{\bf q}\cdot{\bf p}/\mathrm{i}\hbar}.
\end{equation}
The spacing between momentum states is $\Delta_p = 2\pi\hbar/L$,
where $L$ is the edge length of the cubic subsystem,
the volume being $V=L^3$,
and there are $N$ bosons in the subsystem.
Where does this come from?

It is clear that this is an eigenfunction
of the momentum operator,
$\hat {\bf p} = -\mathrm{i}\hbar \nabla_q$.
It is also clear that this is periodic,
$\phi_{\bf p}({\bf q}+L \hat{\bf x}_{j\alpha}) =\phi_{\bf p}({\bf q})$.
But this on its own is a little strange
because the physical system being dealt with
does not consist of a set of periodic replicas of the subsystem.
One would have guessed that the boundary condition
should instead be that the wave function vanished on the boundary
of the subsystem.

The reason for this form is that it ensures that
the momentum operator is Hermitian,
which is a fundamental requirement of quantum mechanics
for an operator that represents a physical observable
(Merzbacher 1970 Messiah  1961).
To see this note that for periodic wave functions,
\begin{eqnarray}
\int_V \mathrm{d}{\bf q}\;
\psi_1({\bf q})^* \hat {\bf p}
\psi_2({\bf q})
& = &
\left.
-\mathrm{i}\hbar \psi_1({\bf q})^* \psi_2({\bf q})
\rule{0cm}{0.4cm}
\right|_{-L/2}^{L/2}
\nonumber \\ && \mbox{ }
+ \mathrm{i}\hbar
\int_V \mathrm{d}{\bf q}\;
\big(\nabla_q \psi_1({\bf q})^*\big) \psi_2({\bf q})
\nonumber \\ & = &
\int_V \mathrm{d}{\bf q}\;
\big(\hat {\bf p}\psi_1({\bf q})\big)^*
\psi_2({\bf q})
\nonumber \\ & \equiv &
\int_V \mathrm{d}{\bf q}\;
\psi_1({\bf q})^*
\hat {\bf p}^\dag \psi_2({\bf q}) .
\end{eqnarray}
The integrated part vanishes on the boundary because of the periodicity,
for example
$\psi(-L/2,y_j,z_j) = \psi(L/2,y_j,z_j)$.
The momentum eigenfunctions above form a complete basis set
for such periodic wave functions.
This links the quantization of the momentum states,
which holds for a finite sized system,
to the Hermitian nature of the momentum operator.

\end{document}